\documentclass[journal,10pt]{IEEEtran}
\usepackage{graphicx}
\usepackage{multirow}
\usepackage{caption}
\usepackage{subcaption}
\usepackage{mathtools}
\usepackage{bm, amsmath, amssymb}
\usepackage{amsmath}
\usepackage{amsfonts}
\usepackage {amssymb,amsmath}
\usepackage{cite}
\usepackage{epstopdf}
\usepackage{microtype}
\DeclareGraphicsExtensions{.pdf,.png,.jpg}

\newcommand {\Define} {\stackrel {\Delta} {=}  }
\newcommand{\mya}{\mathrel{\overset{\makebox[0pt]{{\tiny(a)}}}{=}}}
\newcommand{\myb}{\mathrel{\overset{\makebox[0pt]{{\tiny(b)}}}{=}}}

\newtheorem{theorem}{Theorem}

\usepackage{cite}

\DeclareMathOperator{\sinc}{sinc}

\begin{document}
\title{\vspace{-7mm} Time-Domain to Delay-Doppler Domain Conversion of OTFS Signals in Very High Mobility Scenarios}
\author{\IEEEauthorblockN{Saif Khan Mohammed}
\IEEEauthorblockA{ \thanks{Saif Khan Mohammed is with the Department of Electrical Engineering, Indian Institute of Technology Delhi (IITD), New Delhi, India. Email: saifkmohammed@gmail.com. This work is supported by the Prof. Kishan Gupta and Pramila Gupta Chair at IIT Delhi.}}
}
\maketitle

\vspace{-17mm}
\begin{abstract}
In Orthogonal Time Frequency Space (OTFS) modulation, information symbols are embedded in the delay-Doppler (DD) domain instead of the time-frequency (TF) domain.
In order to ensure compatibility with existing OFDM systems (e.g. 4G LTE), most prior work on OTFS receivers consider a two-step conversion, where the received time-domain (TD) signal is firstly converted to a time-frequency (TF) signal (using an OFDM demodulator) followed by post-processing of this TF signal into a DD domain signal.  
In this paper, we show that the spectral efficiency (SE) performance of a two-step conversion based receiver degrades in very high mobility scenarios where the Doppler shift is a significant fraction of the communication bandwidth (e.g., control and non-payload communication (CNPC) in Unmanned Aircraft Systems (UAS)).  
We therefore consider an {\em alternate} conversion, where the received TD signal is directly converted to the DD domain. The resulting received DD domain signal is shown to be not the same
as that obtained in the two-step conversion considered in prior works.
The alternate conversion does not require an OFDM demodulator
and is shown to have lower complexity than the two-step conversion. 
Analysis and simulations reveal that even in very high mobility scenarios, the SE achieved
with the alternate conversion is {\em invariant} of Doppler shift and is significantly higher than the SE achieved with two-step conversion (which degrades with increasing Doppler shift).                       
\end{abstract}

\section{Introduction}
Recently, Orthogonal Time Frequency Space (OTFS) modulation has been introduced, which has been shown to be more robust towards channel induced
Doppler shift when compared to multi-carrier systems (e.g., Orthogonal Frequency Division Multiplexing (OFDM)) \cite{HadaniOTFS1, HadaniOTFS2, HadaniOTFS3}.
This is because, in OTFS modulation, information is embedded in the delay-Doppler (DD) domain instead of the time-frequency (TF) domain.

Due to the sparseness of the effective channel in the DD domain \cite{HadaniOTFS1, HadaniOTFS2, HadaniOTFS3}, most OTFS receivers first convert the received time-domain (TD) OTFS signal to the DD domain and then perform detection in the DD domain \cite{channel, Emanuele2, mimootfs}.\footnote{\footnotesize{In \cite{lmmse}, a time-domain based Linear Minimum Mean Squared Estimation (LMMSE) detector has been proposed. This detector considers the channel path delay and Doppler shifts to be integer multiples of the delay and Doppler domain resolution which results in a sparse banded structure of the inverse LMMSE equalizer matrix, due to which the equalizer matrix can be computed at low complexity. In the very high Doppler spread scenarios considered here,
the sparse banded structure is not valid and therefore this LMMSE detector will have very high complexity in such scenarios.}}  
In order to ensure compatibility with existing OFDM based receivers, most prior work on OTFS consider a two-step TD to DD domain conversion, where, i) the received TD signal is converted to a TF signal using the OFDM demodulator, and ii) this TF signal is then converted
to the DD domain \cite{HadaniOTFS1, channel, Emanuele2, mimootfs}. Although the two-step conversion based OTFS receiver is significantly more robust to Doppler shift when compared to OFDM \cite{HadaniOTFS1, HadaniOTFS2, HadaniOTFS3}, in this paper we show that for very high mobility scenarios where the Doppler shift is a significant fraction of the communication bandwidth, the performance of the two-step conversion based OTFS receiver {\em degrades} with increasing Doppler shift.\footnote{\footnotesize{In \cite{chocks2020}, a low-complexity TD based OTFS equalizer has been proposed. The effective channel matrix for which equalization is performed, is exactly same as the effective DD domain channel in a two-step receiver, and therefore just like the two-step receiver, the performance of this equalizer will also degrade in very high mobility scenarios.}} 
        
In this paper, we therefore consider an {\em alternate} conversion, where the received TD signal is directly converted to the DD domain using its ZAK representation. There are possibly several variants of the ZAK representation \cite{Zak1}, but in this paper we use the one considered by A. J. E. M. Janssen in \cite{Janssen}. Subsequently, we refer to an OTFS receiver based on this alternate conversion as the ``ZAK receiver" and that based on the two-step conversion as the ``two-step receiver".
The {\em novel} contributions of this paper are:
\begin{itemize}
\item In Section \ref{seczakotfsig} we derive an expression for the ZAK representation of the transmitted TD OTFS signal in terms of the continuous delay and Doppler variables, $\tau$ and $\nu$ respectively.
\item In Section \ref{seczakrx} we consider an alternate TD to DD domain conversion where the ZAK representation of the received TD signal is sampled at discrete
points in the DD domain. The resulting expression for the sampled DD domain signal is not the same
as that obtained in the two-step conversion considered in prior works.
\item We show that the ZAK representation based alternate conversion has a smaller complexity
when compared to the two-step conversion.
\item Through analysis and simulations we show that in {\em very high} mobility scenarios, the spectral efficiency (SE) achieved by the ZAK receiver is almost {\em invariant}
of the Doppler shift and is greater than the SE achieved by the two-step receiver (which degrades with increasing
mobile speed). One such scenario is the control and non-payload communication (CNPC) between
an Unmanned Aircraft System (UAS) and a Ground Station (GS). This involves low bandwidth communication ($30$ KHz) over a wireless channel where the
Doppler shift can be very high due to high UAS speed \cite{ICTCPaper, Haas}.
\end{itemize}    
{\em Notations}: For $x \in {\mathbb R}$, $\lfloor x \rfloor$ is the greatest integer smaller than or equal to $x$.
${\mathbb E}[\cdot]$ denotes the expectation operator. ${\mathcal C}{\mathcal N}(0, \sigma^2)$ denotes
the circular symmetric complex Gaussian distribution with variance $\sigma^2$. For any
matrix ${\bf A}$, $\vert {\bf A} \vert$ denotes its determinant and ${ A}[p,q]$ denotes the
element in its $p$-th row and $q$-th column.          
         
\section{OTFS Modulation}
In OTFS modulation, the delay-Doppler (DD) domain information symbols $x[k,l] \,,\, k=0,1,\cdots, N-1 \,,\, l=0,1,\cdots, M-1$ are firstly converted to Time-Frequency (TF) symbols $X[n,m]\,,\, m=0,1,\cdots, M-1\,,\, n=0,1,\cdots, N-1$
through the inverse Symplectic Finite Fourier Transform (ISFFT) \cite{HadaniOTFS1}, i.e.

{\vspace{-4mm}
\small
\begin{eqnarray}
\label{otfs1}
X[n,m] & \hspace{-3mm} = \hspace{-3mm} & \frac{1}{{MN}}  \sum\limits_{k=0}^{N-1}\sum\limits_{l=0}^{M-1}  x[k,l] \, e^{-j 2 \pi \left(\frac{m l}{M} - \frac{n k}{N}   \right)}.
\end{eqnarray}  
\normalsize}
For a given $T > 0 $ and $\Delta f = 1/T$, these TF symbols are then used to generate the time-domain (TD) transmit signal which is given by the Heisenberg transform \cite{HadaniOTFS1}, i.e.

{\vspace{-4mm}
\small
\begin{eqnarray}
\label{otfs2}
x(t) & = &  \sum\limits_{m=0}^{M-1}\sum\limits_{n=0}^{N-1} X[n,m] \, g(t - nT)  \, e^{j 2 \pi m \Delta f (t - nT)}
\end{eqnarray}
\normalsize}
where $g(\cdot)$ is the transmit pulse. When $g(t)$ is approximately time-limited to $[0 \,,\, T]$, the Heisenberg transform in (\ref{otfs2}) is similar to OFDM
where $X[n,m]$ is the symbol transmitted on the $m$-th sub-carrier ($m=0,1,\cdots, M-1$) of the $n$-th OFDM symbol ($n =0,1,\cdots, N-1$). Each OFDM symbol is of duration $T$ and the sub-carrier spacing is $\Delta f$, i.e., each OTFS frame has duration $NT$ and occupies bandwidth $M \Delta f$.

Let us consider a channel consisting of $L$ paths, where the $i$-th path has gain
$h_i$, induces a delay $\tau_i$ ($0 < \tau_i < T$) and a Doppler shift $\nu_i \,,\, i=1,2,\cdots,L$.
Then, with $x(t)$ as the transmitted signal,
the received signal is given by \cite{Bello}

{\vspace{-4mm}
\small
\begin{eqnarray}
\label{zse1}
y(t) & =  & \sum\limits_{i=1}^L h_i  \, x(t - \tau_i) \, e^{j 2 \pi \nu_i (t - \tau_i)} \, + \, n(t)
\end{eqnarray}
\normalsize}
where $n(t)$ is the additive white Gaussian noise (AWGN) with power spectral density $N_0$.
Most prior work
consider a two-step OTFS receiver which is compatible with OFDM receivers and where the received TD signal $y(t)$ is first converted to a discrete TF signal
through the Wigner transform i.e., ${\Tilde Y}[n,m] \Define \int_{-\infty}^{\infty} g(t - nT) y(t) e^{-j 2 \pi m \Delta f t} dt \,,\, m=0,1,\cdots, M-1, n=0,1,\cdots, N-1$.
This discrete TF signal is then converted to a DD domain signal through the Symplectic Finite Fourier Transform (SFFT) i.e.,
${\widehat x}[k',l'] \Define \sum\limits_{m=0}^{M-1}\sum\limits_{n=0}^{N-1} {\Tilde Y}[n,m] e^{j 2 \pi \left( \frac{m l'}{M} - \frac{n k'}{N}\right)}  \,,\, k'=0,1,\cdots,N-1 , l'=0,1,\cdots,M-1$ \cite{HadaniOTFS1, channel, Emanuele2, mimootfs}. With $g(t)=1/\sqrt{T}, 0 \leq t < T$ and zero otherwise, from (\ref{otfs1}), (\ref{otfs2}), (\ref{zse1}) and the two-step receiver operations (i.e., Wigner transform and SFFT), it follows that

{\vspace{-4mm}
\small
\begin{eqnarray}
\label{xhattwostep}
{\widehat x}[k',l'] & = & \sum\limits_{k=0}^{N-1}\sum\limits_{l=0}^{M-1} x[k,l] \, {\widehat h}[k',l',k,l]  \,\, + \,\, {\widehat n}[k',l']
\end{eqnarray}
\normalsize}
where ${\widehat n}[k',l']$ are the DD domain noise samples and ${\widehat h}[k',l',k,l]$ is given by (\ref{hklotfs}) (see top of next page).   
\begin{figure*}
{\vspace{-11mm}
\small
\begin{eqnarray}
\label{hklotfs}
{\widehat h}[k',l',k,l] & \hspace{-3mm} =  &  \hspace{-3mm} \sum\limits_{i=1}^L h_i e^{-j 2 \pi \frac{\nu_i}{\Delta f} \frac{\tau_i}{T}} \left(  {\Bigg [} \frac{1}{N} \sum\limits_{n=0}^{N-1} \hspace{-1mm} e^{-j 2 \pi n \left( \frac{k' - k}{N} - \frac{\nu_i}{\Delta f}  \right)} {\Bigg ]} {\widehat h}_{i,1}[l',l]   \,  +   \,   e^{j 2 \pi \left(  \frac{\nu_i}{\Delta f}  -  \frac{k'}{N} \right)}   {\Bigg [} \frac{1}{N} \sum\limits_{n=0}^{N-2} \hspace{-1mm} e^{-j 2 \pi n \left( \frac{k' - k}{N} - \frac{\nu_i}{\Delta f}  \right)} {\Bigg ]} {\widehat h}_{i,2}[l',l] \right), \nonumber \\
{\widehat h}_{i,1}[l',l]  & \hspace{-3mm} \Define  &  \hspace{-3mm}   \frac{ \left( 1 - \frac{\tau_i}{T} \right) }{M} \sum\limits_{m=0}^{M-1} \sum\limits_{m'=0}^{M-1}  \hspace{-1mm} e^{j 2 \pi \left(\frac{m'l'}{M}  - \frac{m l}{M} - \frac{m \tau_i}{T} \right)} e^{j \pi \left( 1 + \frac{\tau_i}{T}  \right) \left( \frac{\nu_i}{\Delta f} - (m' - m) \right)}   \sinc\left( \left(1 - \frac{\tau_i}{T} \right) \left(\frac{\nu_i}{\Delta f} - (m' - m) \right) \right), \nonumber \\
{\widehat h}_{i,2}[l',l]  & \hspace{-3mm}  \Define  &  \hspace{-3mm}   \frac{ \left( \frac{\tau_i}{T} \right) }{M} \sum\limits_{m=0}^{M-1} \sum\limits_{m'=0}^{M-1}  \hspace{-1mm} e^{j 2 \pi \left(\frac{m'l'}{M}  - \frac{m l}{M} - \frac{m \tau_i}{T} \right)} e^{j \pi \frac{\tau_i}{T} \left( \frac{\nu_i}{\Delta f} - (m' - m) \right)}   \sinc\left( \frac{\tau_i}{T}  \left(\frac{\nu_i}{\Delta f} - (m' - m) \right) \right).
\end{eqnarray}
\normalsize}
\end{figure*}
 
\section{The ZAK Representation of OTFS Signal}
\label{seczakotfsig}
The ZAK transform of $x(t)$ is given by \cite{Janssen}

{\vspace{-4mm}
\small
\begin{eqnarray}
\label{zdef1}
{\mathcal Z}_x(\tau, \nu) & \Define & \sqrt{T} \, \sum\limits_{k=-\infty}^{\infty} \, x(\tau + kT) \, e^{-j 2 \pi k \nu T } \,,\, \nonumber \\
& & -\infty < \tau < \infty \,,\, -\infty < \nu < \infty. 
\end{eqnarray}
\normalsize}
Due to the following result, the $\tau-$ and $\nu-$ domains are called
the ``delay" and ``Doppler" domain respectively, i.e., collectively we refer to them as the delay-Doppler (DD) domain. 
\begin{theorem}
\label{zlem52}
Let $x(t)$ be the transmitted signal and let there be only one channel path with
a delay of $\tau_0$ and a Doppler shift of $\nu_0$. The ZAK transform of the noise-free received signal
$y(t) = x(t - \tau_0) e^{j 2 \pi \nu_0 (t - \tau_0)}$ is then given by
\begin{eqnarray}
\label{zakmod2}
{\mathcal Z}_y(\tau, \nu) & = & e^{j 2 \pi \nu_0 (\tau - \tau_0)} {\mathcal Z}_x(\tau - \tau_0 , \nu - \nu_0)
\end{eqnarray}
i.e., delay and Doppler shift in time-domain results in a shift along the $\tau-$ and $\nu-$ domains by
$\tau_0$ and $\nu_0$ respectively. 
\end{theorem}
\begin{IEEEproof}
From (\ref{zdef1}), the ZAK transform of $y(t)$ is given by
{\vspace{-2mm}
\small
\begin{eqnarray}
\label{zakmod1}
{\mathcal Z}_y(\tau , \nu) & \hspace{-3mm}  = & \hspace{-3mm} \sqrt{T} \sum\limits_{k=-\infty}^{\infty} y(\tau + kT) \, e^{-j 2 \pi \nu k T} \nonumber \\
& \hspace{-8mm} = & \hspace{-5mm} \sqrt{T} \sum\limits_{k=-\infty}^{\infty}  \hspace{-1mm} x(\tau - \tau_0 + kT) \, e^{j 2 \pi \nu_0 (\tau + kT - \tau_0)} \, e^{-j 2 \pi \nu k T} \nonumber \\
& \hspace{-8mm}  = & \hspace{-5mm} e^{j 2 \pi \nu_0 (\tau - \tau_0)} \sqrt{T} \sum\limits_{k=-\infty}^{\infty} \hspace{-1mm}  x(\tau - \tau_0 + kT) \, e^{-j 2 \pi (\nu - \nu_0) k T} \nonumber \\
& \hspace{-8mm} \mya & \hspace{-5mm} e^{j 2 \pi \nu_0 (\tau - \tau_0)} {\mathcal Z}_x(\tau - \tau_0 , \nu - \nu_0)
\end{eqnarray}
\normalsize}
where step (a) follows from (\ref{zdef1}). 
\end{IEEEproof}
Let us consider $g(t)$ to be

{\vspace{-4mm}
\small
\begin{eqnarray}
\label{zprf7}
g(t) & = 
\begin{cases}
\frac{1}{\sqrt{T}} &, 0 \leq t < T \\
0 &, \mbox{\small{otherwise}} \\
\end{cases}.
\end{eqnarray} 
\normalsize}
From (\ref{zdef1}) it follows that in the interval ${\mathcal S} \Define \{ (\tau,\nu) \, | \, 0 \leq \tau < T \,,\, 0 \leq \nu < \Delta f \}$, the ZAK transform of $g(t)$ in (\ref{zprf7}) is

{\vspace{-4mm}
\small
\begin{eqnarray}
\label{zprf88}
{\mathcal Z}_g(\tau , \nu) & = & 1 \,,\, 0 \leq \tau < T \,,\, 0 \leq \nu < \Delta f.
\end{eqnarray}
\normalsize}
From (\ref{zdef1}) it follows that for any $n \in {\mathbb Z}$

{\vspace{-4mm}
\small
\begin{eqnarray}
\label{zprf89}
{\mathcal Z}_g(\tau - nT , \nu) & \hspace{-3mm}  = &  \hspace{-3mm} \sqrt{T} \sum\limits_{k=-\infty}^{\infty} \hspace{-2mm} g(\tau - nT + kT) \, e^{-j 2 \pi k \nu T} \nonumber \\
& \hspace{-25mm}  = & \hspace{-14mm} e^{-j 2 \pi \nu n T} \sqrt{T} \hspace{-1mm} \sum\limits_{k=-\infty}^{\infty} \hspace{-2mm} g(\tau +(k - n)T) \, e^{-j 2 \pi (k-n) \nu T} \nonumber \\
& \hspace{-25mm} =  & \hspace{-14mm}  e^{-j 2 \pi \nu n T} \, {\mathcal Z}_g(\tau, \nu)
\end{eqnarray}
\normalsize}
i.e., ${\mathcal Z}_g(\tau,\nu)$ is {\em quasi-periodic} along the delay domain \cite{Janssen}.
From the above derivation it is clear that quasi-periodicity
holds not just for ${\mathcal Z}_g(\tau,\nu)$ but is valid for the ZAK transform of any finite energy time-domain signal. Also,
for any $m \in {\mathbb Z}$, ${\mathcal Z}_g(\tau , \nu + m\Delta f) = {\mathcal Z}_g(\tau , \nu)$ (see $(2.20)$ and $(2.21)$ in \cite{Janssen}).
Using these facts along with (\ref{zprf88}) it follows that
\begin{eqnarray}
\label{zprf8}
\hspace{-3mm} {\mathcal Z}_g(\tau , \nu) & \hspace{-3mm}  = &  \hspace{-3mm} e^{j 2 \pi \nu \left\lfloor \frac{\tau}{T}\right\rfloor T} \,,\, -\infty < \tau < \infty \,,\, -\infty < \nu < \infty.
\end{eqnarray}
The following theorem gives the expression for the ZAK transform of the transmit OTFS signal $x(t)$.
\begin{theorem}
\label{zthm2}
The ZAK transform of $x(t)$ in (\ref{otfs2}) is

{\vspace{-4mm}
\small
\begin{eqnarray}
\label{zprf6}
\hspace{-4mm} {\mathcal Z}_x(\tau, \nu) & \hspace{-3mm}  = & \hspace{-3mm} \sum\limits_{k=0}^{N-1}\sum\limits_{l=0}^{M-1} x[k,l] \, \Phi_{k,l}(\tau,\nu) \,,\,  \nonumber \\ 
\hspace{-4mm} \Phi_{k,l}(\tau,\nu) & \hspace{-3mm}  \Define & \hspace{-3mm} \frac{{\mathcal Z}_g(\tau,\nu)}{MN}  \sum\limits_{m=0}^{M-1}\sum\limits_{n=0}^{N-1} {\Bigg [}  e^{-j 2 \pi n T \left( \nu - k \frac{\Delta f}{N} \right)} \nonumber \\
& & \hspace{30mm}  e^{j 2 \pi m \Delta f \left( \tau - \frac{lT}{M} \right)} {\Bigg]}, 
\end{eqnarray}
where ${\mathcal Z}_g(\tau,\nu)$ is given by (\ref{zprf8}).
\normalsize}
\end{theorem}
\begin{IEEEproof}
See Appendix \ref{appendix_thm2}.
\end{IEEEproof}
From Theorem \ref{zthm2}, it is clear that $\Phi_{k,l}(\tau,\nu)$ are the basis signals in the DD domain.
Taking the expression of ${\mathcal Z}_g(\tau , \nu)$ from (\ref{zprf8}) into (\ref{zprf6}), and using the fact that $T \Delta f = 1$, we get

{
\vspace{-4mm}
\small
\begin{eqnarray}
\label{zprf9}
\hspace{-3mm} \Phi_{k,l}(\tau, \nu) & \hspace{-3mm} = &  \hspace{-3mm} e^{j 2 \pi \frac{\nu}{\Delta f}  \left\lfloor \frac{\tau}{T}\right\rfloor } w_{1}\left(\nu - \frac{k \Delta f}{N}\right) \, w_{2}\left(\tau - \frac{lT}{M}\right) \,,\,  \nonumber \\
\hspace{-4mm} w_{1}(\nu)  & \hspace{-3.5mm}  \Define  &  \hspace{-3.5mm}  \frac{1}{N} \hspace{-1mm} \sum\limits_{n=0}^{N-1}  \hspace{-1mm} e^{-j 2 \pi n \nu T }   =  e^{-j \pi (N-1) \frac{\nu}{\Delta f}}  \frac{\sin\left( \pi N \frac{\nu}{\Delta f} \right)}{N\sin\left( \pi \frac{\nu}{ \Delta f} \right)} \nonumber \\
\hspace{-4mm} w_{2}(\tau)  & \hspace{-3.5mm}  \Define & \hspace{-3.5mm}  \frac{1}{M} \hspace{-1mm} \sum\limits_{m=0}^{M-1}  \hspace{-1mm} e^{j 2 \pi m \Delta f \tau }  =  e^{j \pi (M-1) \frac{\tau}{T} } \frac{\sin\left( \pi M  \frac{\tau}{T}  \right)}{M\sin\left(\pi \frac{\tau}{T} \right)}.
\end{eqnarray}
\normalsize}
\section{An Alternate TD to DD Domain Conversion}
\label{seczakrx}
In (\ref{zprf9}), apart from the quasi-periodicity term $e^{j 2 \pi \frac{\nu}{\Delta f} \left\lfloor \frac{\tau}{T}\right\rfloor }$, the $(k,l)$-th basis signal $\Phi_{k,l}(\tau,\nu)$ is a {\em product} of two signals, $w_{1}\left(\nu - \frac{k \Delta f}{N}\right)$ and $w_{2}\left(\tau - \frac{lT}{M}\right)$, which are shifted versions of the basic pulses $w_1(\nu)$ and $w_2(\tau)$ along the Doppler and delay domain respectively.
From (\ref{zprf9}) it is clear that $w_{2}\left(\tau - \frac{lT}{M}\right)$ and $w_{1}\left(\nu - \frac{k \Delta f}{N}\right)$ are localized around $\tau =  l T/M$ along the delay domain and around $\nu = k \Delta f/N$ along the Doppler domain, respectively.
From (\ref{zprf9}) it is also clear that the basic pulses have a null at the centre location of all other shifted versions, i.e., $w_{1}(\nu = k \Delta f/N) = 0 \,,\, k = 1,\cdots, N-1$ and
$w_{2}(\tau = l T/M) = 0 \,,\, l = 1,\cdots, M-1$. This suggests a receiver where the received TD signal $y(t)$ is directly transformed to its ZAK representation ${\mathcal Z}_y(\tau, \nu)$, followed by
sampling ${\mathcal Z}_y(\tau, \nu)$
at integer multiples of $T/M$ along the delay domain and at integer multiples of $\Delta f/N$ along the Doppler domain.

In this paper, we therefore consider an alternate conversion of the TD signal $y(t)$ in (\ref{zse1}) to a DD domain signal, by sampling the ZAK transform of $y(t)$ (i.e., ${\mathcal Z}_y(\tau,\nu)$) at discrete points $(\tau = l'T/M, \nu = k'\Delta f/N)$, $l'=0,1,\cdots,M-1, k'=0,1,\cdots,N-1,$ in the DD domain. 
Subsequently, we refer to a OTFS receiver using this alternate conversion, as the ``ZAK receiver".
The sampled DD domain signal in the ZAK receiver is then given by

{\vspace{-4mm}
\small
\begin{eqnarray}
\label{zprf11}
Y[k',l'] & \Define & {\mathcal Z}_y\left(\tau = \frac{l'T}{M} \,,\, \nu = \frac{k' \Delta f}{N} \right) \,,\, \nonumber \\
& \mya &   \sqrt{T} \sum\limits_{n=0}^{N} y\left(nT + \frac{l'T}{M}  \right) e^{-j 2 \pi n \frac{k'}{N}} \nonumber \\
\hspace{-12mm} & \hspace{-8mm}  & \hspace{-4mm} k' = 0,1,\cdots, N-1 \,,\, l' = 0,1,\cdots, M-1 
\end{eqnarray}  
\normalsize}
where step (a) follows from (\ref{zdef1}) and the fact that the received signal $y(t)$ is time-limited to the interval $0 < t < (N+1)T$. This is because, the channel path delays are less than $T$ ($0 < \tau_i < T, i=1,2,\cdots,L$) and the transmit signal $x(t)$ is limited to the time interval $0 \leq t < NT$ as $g(t)$ is time-limited
to $0 \leq t < T$ (see the expression for $x(t)$ in (\ref{otfs2})). We also note that the discrete time-domain samples $y\left(nT + \frac{l'T}{M}  \right) \,,\, n=0,1,\cdots, N, l'=0,1,\cdots, M-1$
are obtained by sampling $y(t)$ at integer multiples of $1/(M \Delta f) = T/M$.  
The ZAK transform of $y(t)$ in (\ref{zse1}) is given by
 
{\vspace{-4mm}
\small
\begin{eqnarray}
\label{zprf12}
{\mathcal Z}_y(\tau,\nu) &  \hspace{-2mm}  \mya &  \hspace{-2mm} \sum\limits_{i=1}^L h_i  \, e^{j 2 \pi \nu_i (\tau - \tau_i)} {\mathcal Z}_x(\tau - \tau_i,\nu - \nu_i)  \, + \,  {\mathcal Z}_{n}(\tau,\nu) \nonumber \\
& \hspace{-25mm}  \myb & \hspace{-15mm} \sum\limits_{k=0}^{N-1}\sum\limits_{l=0}^{M-1} \hspace{-1mm} x[k,l] {\Bigg [} \sum\limits_{i=1}^L  h_i  \, e^{j 2 \pi \nu_i (\tau - \tau_i)} e^{j 2 \pi \frac{(\nu - \nu_i)}{\Delta f} \left\lfloor \frac{\tau - \tau_i}{T}  \right\rfloor }  \nonumber \\
& & \hspace{-10mm}   w_{1}\hspace{-1mm}\left(\nu - \nu_i - \frac{k \Delta f}{N} \right) w_{2}\hspace{-1mm}\left(\tau - \tau_i - \frac{lT}{M}\right)  {\Bigg ]} + {\mathcal Z}_{n}(\tau,\nu)
\end{eqnarray}
\normalsize}
where step (a) follows from Theorem \ref{zlem52} and step (b) follows from Theorem \ref{zthm2} and (\ref{zprf9}).
In (\ref{zprf12}), ${\mathcal Z}_{n}(\tau,\nu)$ is the ZAK transform of the AWGN $n(t)$.
Using (\ref{zprf12}) in (\ref{zprf11}), we get

{\vspace{-4mm}
\small
\begin{eqnarray}
\label{zprf13}
\hspace{-1mm} Y[k',l'] & \hspace{-4.5mm} = &  \hspace{-4.5mm}   \sum\limits_{k=0}^{N-1}\sum\limits_{l=0}^{M-1}  x[k,l] \, {\Tilde h}[k',l',k,l]  \,  + \,  Z[k',l'] \nonumber \\
{\Tilde h}[k',l',k,l]  & \Define & \sum\limits_{i=1}^L  h_i[k',l']  \,  {\Bigg [} w_{1}\hspace{-1mm}\left(\frac{(k' - k) \Delta f}{N} - \nu_i  \right) \nonumber \\
& & \hspace{21mm} w_{2}\hspace{-1mm}\left( \frac{(l' - l)T}{M} - \tau_i \right)  {\Bigg ]} \nonumber \\
Z[k',l'] & \Define & {\mathcal Z}_{n}\left({l'T}/{M},{k' \Delta f}/{N}\right) \nonumber \\
h_i[k',l'] & \hspace{-3mm}  \Define &  \hspace{-3mm} h_i  \, e^{j 2 \pi \frac{\nu_i}{\Delta f} \left(\frac{l'}{M} - \frac{\tau_i}{T}\right)} e^{j 2 \pi \left(\frac{k' }{N} -  \frac{\nu_i}{\Delta f} \right) \left\lfloor \left(\frac{l'}{M} - \frac{\tau_i}{T}\right)  \right\rfloor } \nonumber \\
& & \hspace{-8mm} k'=0,1,\cdots, N -1 \,,\, l' =0,1,\cdots, M-1.
\end{eqnarray}
\normalsize}
The noise samples in the DD domain are given by

{\vspace{-4mm}
\small
\begin{eqnarray}
\label{zprf73}
\hspace{-1mm} Z[k',l'] & \hspace{-3mm}  = & \hspace{-3mm} {\mathcal Z}_{n}\left(\frac{l'T}{M},\frac{k' \Delta f}{N}\right)  \mya  \sqrt{T} \sum\limits_{k=0}^{N}  n\left(\frac{l' T}{M} + kT\right) e^{-j 2 \pi k \frac{k'}{N}} \nonumber \\
& \hspace{-12mm} = &  \hspace{-8mm} \sqrt{T} \sum\limits_{k=0}^{N-1} n\left(\frac{l' T}{M} + kT\right) e^{-j 2 \pi k \frac{k'}{N}} \, + \, \sqrt{T} \, n \left(\frac{l'T}{M} + NT  \right)\nonumber \\
& & k'=0,1,\cdots, N -1 \,,\, l' =0,1,\cdots, M-1
\end{eqnarray}
\normalsize}
where step (a) follows from (\ref{zdef1}) and the fact that the received signal $y(t)$ is time-limited to $0 < t < NT$.
By comparing the expression for ${\widehat h}[k',l',k,l]$ in (\ref{hklotfs}) and ${\Tilde h}[k',l',k,l]$ in (\ref{zprf13}) it is clear that the received discrete DD domain signal
for the two-step receiver (i.e. ${\widehat x}[k',l']$ in (\ref{xhattwostep})) is not the same as that for the ZAK receiver (see $Y[k',l']$ in (\ref{zprf13})).    

From (\ref{zprf11}) it is clear that for a given $l'$, $Y[k',l']$
can be computed for all $k' = 0, 1, \cdots, N-1$ using a N-point Discrete Fourier Transform (DFT) whose complexity is $O(N \log N)$. Therefore the complexity of the alternate conversion, i.e., computing
$Y[k',l']$ from $y(t)$ for all $k'=0,1,\cdots, N-1 \,,\, l'=0,1,\cdots, M-1$ is $O(MN \log N)$.
For the two-step conversion, the complexity of computing SFFT is $O(MN \log (MN))$ which is greater than $O(MN\log (N))$.
Further, the alternate conversion does not require an OFDM demodulator.

With information symbols $x[k,l] \sim \, \mbox{i.i.d.} \, {\mathcal C}{\mathcal N}(0, E)$, ${\mathbb E}\left[ \int_{-\infty}^{\infty} \hspace{-1mm} \vert x(t) \vert^2 dt \right]   =  E$
(follows from (\ref{zprf61}) in Appendix \ref{appendix_thm2}). 
Since $x(t)$ is time-limited to $NT$ seconds, the average transmit power is $E/NT$. With a communication bandwidth of
$M \Delta f$, the AWGN power at the receiver is $M \Delta f N_0$ and therefore the ratio of the transmit power to the receiver noise
power is

{\vspace{-4mm}
\small
\begin{eqnarray}
\label{eqnrho}
\rho & \Define & \frac{E/NT}{M \Delta f N_0} = \frac{E}{MN N_0}.
\end{eqnarray} 
\normalsize}

Let the received DD domain samples $Y[k',l']$ and the transmitted information symbols $x[k,l]$ be arranged into vectors ${\bf y}$ and ${\bf x}$ 
respectively as given by (\ref{zprf14}) (see top of next page).
\begin{figure*}
{\vspace{-11mm}
\small
\begin{eqnarray}
\label{zprf14}
{\bf y}  &  \Define  & (Y[0,0], Y[0,1], \cdots, Y[0, M-1], Y[1,0], \cdots, Y[1,M-1], \cdots, Y[N-1, 0], \cdots, Y[N-1,M-1])^T \nonumber \\
{\bf x}  &  \Define  & (x[0,0], x[0,1], \cdots, x[0, M-1], x[1,0], \cdots, x[1,M-1], \cdots, x[N-1, 0], \cdots, x[N-1,M-1])^T \nonumber \\
{\bf z}  &  \Define  & (Z[0,0], Z[0,1], \cdots, Z[0, M-1], Z[1,0], \cdots, Z[1,M-1], \cdots, Z[N-1, 0], \cdots, Z[N-1,M-1])^T.
\end{eqnarray}
\normalsize}
\end{figure*}
From (\ref{zprf13}) it then follows that
\begin{eqnarray}
\label{zprf15}
{\bf y} & = & {\bf H} \, {\bf x} \, + \, {\bf z}
\end{eqnarray}
where ${\bf z}$ is the vector of noise samples in the DD domain (see (\ref{zprf14})) and ${\bf H} \in {\mathbb C}^{MN \times MN}$
is the equivalent DD domain channel matrix. The element of ${\bf H}$ in its $(l' + k'M +1)$-th row and $(l + kM +1)$-th column is
${\Tilde h}[k',l',k,l]$ (see (\ref{zprf13})). Let ${\bf K}_{{\bf z}} \Define {\mathbb E}[{\bf z} \, {\bf z}^H]$ denote the covariance matrix of ${\bf z}$. In (\ref{zprf73}), the AWGN samples $n\left(\frac{l' T}{M} + kT\right) \,,\, l' = 0,1,\cdots, M-1, k =0,1,\cdots, N$ are i.i.d. ${\mathcal C}{\mathcal N}(0, M \Delta f N_0)$,\footnote{\footnotesize{As the communication bandwidth is $M \Delta f = M/T$, the received time-domain signal is
filtered with a low-pass filter having bandwidth $M/T$ due to which noise samples taken at integer multiples of $T/M$ are i.i.d. ${\mathcal C}{\mathcal N}(0, M \Delta f N_0)$.}} and therefore the element of ${\bf K}_{{\bf z}}$ in its $(k_1M + l_1 +1)$-th row and
$(k_2M + l_2 +1)$-th column is given by 

{\vspace{-4mm}
\small
\begin{eqnarray}
\label{kzeqn}
{K}_{{\bf z}}[{k_1M + l_1 +1,k_2M + l_2 +1}] & =  {\mathbb E}[ Z[k_1,l_1] Z^*[k_2, l_2]] \nonumber \\
& \hspace{-30mm} = \begin{cases}
M N N_0 \left(1 + \frac{1}{N} \right) &, k_1 = k_2 \,,\, l_1 = l_2 \\
M N N_0  \left( \frac{1}{N} \right) &, k_1 \ne k_2 \,,\, l_1 = l_2 \\
0  &, l_1 \ne l_2 \\
\end{cases}.
\end{eqnarray}
\normalsize}
From (\ref{kzeqn}) it is also clear that
\begin{eqnarray}
\label{largeNeqn}
\lim\limits_{N \rightarrow \infty} \frac{{\bf K}_{{\bf z}}}{MNN_0} & = & {\bf I}
\end{eqnarray}
where ${\bf I}$ denotes the $MN \times MN$ identity matrix. 
With perfect knowledge of ${\bf H}$ at the receiver and joint equalization of all $MN$ information symbols in the DD domain, the achievable spectral efficiency (SE) of the ZAK receiver is given by \cite{DTse}

{\vspace{-4mm}
\small
\begin{eqnarray}
\label{zprfcap}
C_{\mbox{\tiny{Zak}}}  & \mya & \frac{1}{MN} \log_2 \left\vert {\bf I} + \rho \, {\bf H}^{\bf H} \left(\frac{{\bf K}_{{\bf z}}}{MNN_0}\right)^{-1} {\bf H} \right\vert \nonumber \\
& \approx & \frac{1}{MN} \log_2 \left\vert {\bf I} + \rho \, {\bf H}^{\bf H} {\bf H} \right\vert  \,\,\,\, (\mbox{\small{Large $N$}})
\end{eqnarray}
\normalsize}
where step (a) follows from (\ref{eqnrho}) and (\ref{zprf15}), and the approximation follows from (\ref{largeNeqn}).      

{\em Example}: In the following we compare the ZAK receiver and the two-step receiver for a
single-path channel (i.e., $L=1$) with zero delay and a Doppler shift equal to an integer multiple of $\Delta f$ i.e., $\nu_1 = q \Delta f, q \in \{ 0,1,\cdots, M-1\}$.
From (\ref{hklotfs}) and (\ref{zprf13}) we then have

{\vspace{-4mm}
\small
\begin{eqnarray}
\label{cmpotfs}
{\widehat h}[k',l',k,l]  & \hspace{-3mm}  = &  \hspace{-3mm} h_1 e^{j 2 \pi \frac{ql'}{M}} \delta[k - k'] \,  {\Bigg [}  \frac{1}{M} \hspace{-1mm} \sum\limits_{m=0}^{M-1-q} e^{j 2 \pi m \frac{(l' -l)}{M}}  {\Bigg ]}  \nonumber \\
{\Tilde h}[k',l',k,l]  & \hspace{-3mm} = & \hspace{-3mm}  h_1 e^{j 2 \pi \frac{ql'}{M}} \delta[k - k'] \,   \delta[l - l']
\end{eqnarray}  
\normalsize}
where $\delta[k], k \in {\mathbb Z}$ is one when $k=0$ and is zero for all other $k \ne 0$. Just as in (\ref{zprf15}) for the ZAK receiver, for the two-step receiver also, let ${\widehat {\bf H}} \in {\mathcal C}^{MN \times MN}$ denote the effective DD domain channel matrix whose element in its $(k'M + l'+1)$-th row and $(kM + l +1)$-th column is ${\widehat h}[k',l',k,l]$.  
The spectral efficiency achieved by the two-step receiver is $C \Define \left(\log_2 \left\vert {\bf I} + \rho {\widehat {\bf H}}^H {\widehat {\bf H}}  \right\vert\right)/(MN)$. Let ${\widehat I}[k_1, k_2, l_1, l_2]$ denote the element of ${\widehat {\bf H}}^H{\widehat {\bf H}}$ in its $(k_1M + l_1 + 1)$-th row and $(k_2M + l_2 + 1)$-th column. From (\ref{cmpotfs}) it then follows that
\begin{eqnarray}
\label{cmpotfs1}
\hspace{-1mm} {\widehat I}[k_1, k_2, l_1, l_2] & \hspace{-3mm}  = & \hspace{-3mm} \delta[ k_1 - k_2] \vert h_1 \vert^2 \, \frac{1}{M} \hspace{-1mm} \sum\limits_{m=0}^{M - q - 1} \hspace{-2mm} e^{j 2 \pi \frac{m (l_1 - l_2)}{M}}. 
\end{eqnarray}      
For the ZAK receiver, from (\ref{cmpotfs}) it follows that ${\bf H}^H{\bf H} = {\bf H}{\bf H}^H = \vert h_1 \vert^2 {\bf I}$, and therefore from (\ref{zprfcap}) it follows that for large $N$, $C_{\mbox{\tiny{Zak}}} \approx \log_2(1 + \rho \vert h_1 \vert^2)$ which is {\em same} as the capacity of an ideal single-path channel without any Doppler shift. Further, from (\ref{cmpotfs1})
it follows that ${\widehat {\bf H}}^H {\widehat {\bf H}}$ is a block diagonal matrix where each block along the diagonal is a $M \times M$ matrix ${\bf A}$ whose
element in the $(l_1+1)$-th row and $(l_2+1)$-th column is $\vert h_1 \vert^2 \, \frac{1}{M} \hspace{-1mm} \sum\limits_{m=0}^{M - q - 1} \hspace{-2mm} e^{j 2 \pi \frac{m (l_1 - l_2)}{M}}, l_1=0,1,\cdots,M-1\,,\, l_2=0,1,\cdots,M-1$. The spectral efficiency achieved by the two-step receiver is therefore $C = \frac{1}{M} \log_2 \left\vert  {\bf I} + \rho {\bf A} \right\vert$. One can check that ${\bf A}$ is a positive semi-definite matrix and the DFT vectors ${\bf u}_p =  \frac{1}{\sqrt{M}}\left( 1, e^{\frac{j 2 \pi p}{M}}, \cdots, e^{\frac{j 2 \pi p (M-1)}{M}}  \right)^T, p=0,1,\cdots, M-1$ are a set of $M$ orthonormal eigen-vectors of ${\bf A}$ with corresponding eigen-values
$\lambda_p = \vert h_1 \vert^2 \,,\, p=0,1,\cdots,(M-q-1)$ and $\lambda_p = 0 \,,\, p=(M-q), \cdots, M-1$. From this it follows that $C = \frac{1}{M} \log_2 \left\vert  {\bf I} + \rho {\bf A} \right\vert = \left(1 - \frac{q}{M} \right) \log_2(1 + \rho \vert h_1 \vert^2) = \left(1 - \frac{\nu_1}{M \Delta f} \right) \log_2(1 + \rho \vert h_1 \vert^2)$ which is clearly less than $C_{\mbox{\tiny{Zak}}}$.
Therefore, if the ratio of the Doppler shift to the communication bandwidth (i.e., $\nu_1/(M \Delta f)$) is high, the spectral efficiency of the two-step receiver
will be {\em significantly} less than $C_{\mbox{\tiny{Zak}}}$.  

\begin{figure}[t]
\vspace{-0.7 cm}
\hspace{-0.2 in}
\centering
\includegraphics[width= 3.0 in, height= 2.2 in]{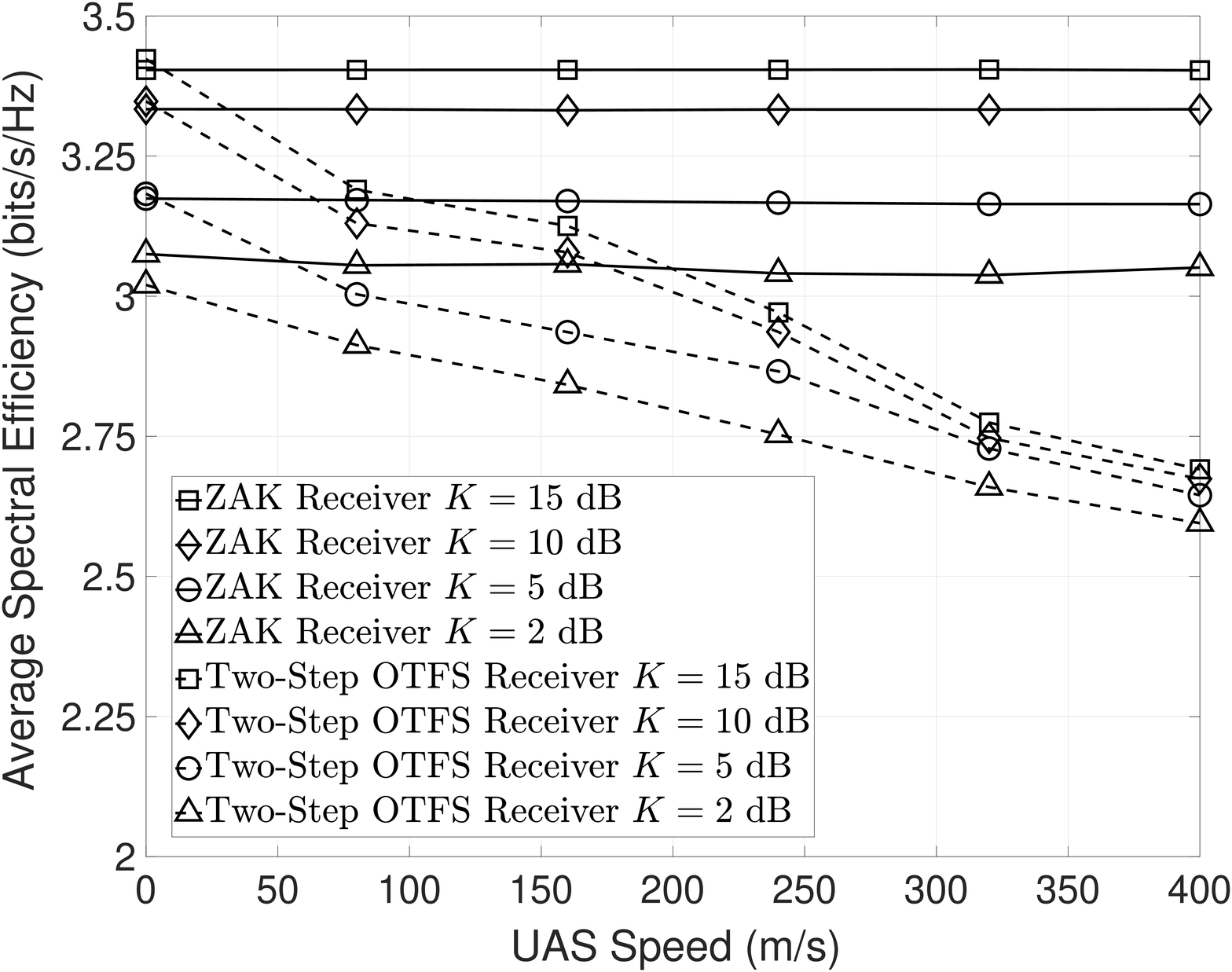}
\vspace{-0.2 cm}
\caption{Spectral Efficiency versus UAS speed (m/s).} 
\vspace{-0.2cm}
\label{Zakrxfig1}
\end{figure}

\begin{figure}[t]
\vspace{-0.2 cm}
\hspace{-0.2 in}
\centering
\includegraphics[width= 3.0 in, height= 2.2 in]{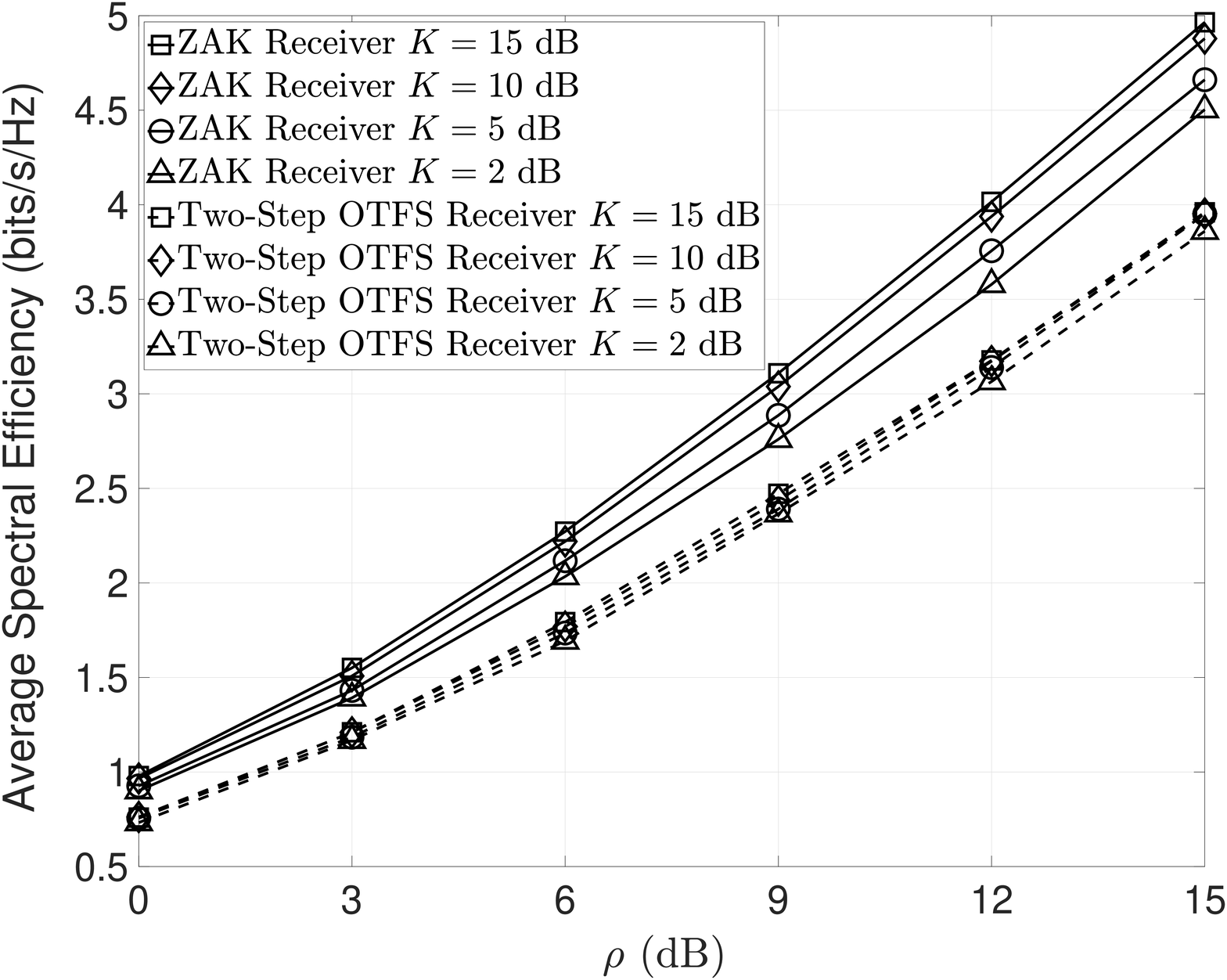}
\vspace{-0.2 cm}
\caption{Spectral Efficiency (bps/Hz) versus $\rho$ (dB).} 
\vspace{-0.4cm}
\label{Zakrxfig2}
\end{figure}

\section{Numerical Simulations}
In this section we compare the average spectral efficiency (SE) achieved by the two-step OTFS receiver and the
ZAK receiver for control and non-payload communication (CNPC) between an Unmanned Aircraft System (UAS)
and a Ground Station (GS) when the aircraft is en-route at an altitude of about $10$ Km \cite{ICTCPaper}. For the en-route scenario a widely
accepted model is the two-path model, with the direct path having a deterministic gain and the reflected path having a Rayleigh faded gain, i.e., $h_1 = \sqrt{K/(K+1)}$, $h_2 \sim {\mathcal C}{\mathcal N}(0, 1/(K+1))$  
\cite{Haas}. In \cite{Haas} it is mentioned that $K$ is typically $15$ dB and is usually not below $2$ dB. The delay between these two paths
is $33 \, \mu s$ (i.e., $\tau_1=0, \tau_2 = 33 \, \mu s$).
The Doppler shift for the two paths is $\nu_1 = v f_c/c, \nu_2 = (v f_c/c) \cos\left(\pi - \omega U \right)$ where $v$ is the speed of the aircraft, $c$ is the speed of light, $f_c$ is the carrier frequency,
$U$ is uniformly distributed in $[0 \,,\, 1]$ and $\omega = 3.5^{\circ}$ is the Doppler beamwidth \cite{Haas}.
The total time duration for a downlink CNPC frame is $23$ ms, the channel bandwidth is $30$ KHz and $f_c = 5.06$ GHz \cite{ICTCPaper}. We therefore choose $M = 15, N = 46$ and $\Delta f = 2 $ KHz.

In Fig.~\ref{Zakrxfig1} we plot the average SE achieved by the two-step receiver (i.e., ${\mathbb E}[C] = (1/MN){\mathbb E}\left[\log_2 \left\vert {\bf I} + \rho {\widehat {\bf H}}^H {\widehat {\bf H}}  \right\vert \right]$) and that achieved by the ZAK receiver (i.e., ${\mathbb E}[C_{\mbox{\tiny{Zak}}}]$, see (\ref{zprfcap})), as a function of increasing UAS speed for a fixed $\rho = 10$ dB. The averaging of the SE is w.r.t. $h_2$ and $\nu_2$. It is observed that the
SE of the two-step receiver decreases with increasing UAS speed whereas that of the ZAK receiver is almost {\em invariant} of the Doppler shift.
At high UAS speed of $400$ m/s (ratio of the maximum Doppler shift to the bandwidth is $0.225$) and a typical $K=15$ dB, the SE achieved by the ZAK receiver is roughly $0.7$ bits/s/Hz higher than that achieved by a two-step receiver.
In Fig.~\ref{Zakrxfig2}, we study the SE comparison as a function of increasing $\rho$ for a fixed UAS speed of $400$ m/s. To achieve a desired high SE, the two-step receiver needs a significantly higher transmit power when compared to
the ZAK receiver.

\appendices

\section{Proof of Theorem \ref{zthm2}}
\label{appendix_thm2}
From (\ref{otfs1}) and (\ref{otfs2}) it follows that 

{\vspace{-4mm}
\small
\begin{eqnarray}
\label{zprf61}
x(t) & \hspace{-2mm}  = & \hspace{-2mm} \sum\limits_{k=0}^{N-1}\sum\limits_{l=0}^{M-1} x[k,l] \, \phi_{k,l}(t) \nonumber \\
\phi_{k,l}(t) & \hspace{-2mm} \Define &  \hspace{-2mm} \sum\limits_{m=0}^{M-1}\sum\limits_{n=0}^{N-1} \hspace{-1mm} \frac{g(t - nT)}{MN}  e^{j 2 \pi n \frac{k}{N}} e^{j 2 \pi m \Delta f \left(t - \frac{lT}{M} \right)}.
\end{eqnarray}
\normalsize}
Taking the ZAK transform of the expression for $x(t)$ in (\ref{zprf61}), from (\ref{zdef1}) we have

{\vspace{-4mm}
\small
\begin{eqnarray}
\label{zprf62}
{\mathcal Z}_x(\tau,\nu) & = & \sum\limits_{k=0}^{N-1}\sum\limits_{l=0}^{M-1} x[k,l] \, \Phi_{k,l}(\tau,\nu) \,,\, \mbox{\small{where}}\nonumber \\
 \Phi_{k,l}(\tau,\nu) & \mya & \sqrt{T} \sum\limits_{k' = -\infty}^{\infty} \phi_{k,l}(\tau+k' T) e^{-j 2 \pi \nu k' T} \nonumber \\
& \hspace{-39mm}  \myb &  \hspace{-22mm} \sum\limits_{m=0}^{M-1}\sum\limits_{n=0}^{N-1} \hspace{-1mm} \frac{1}{MN} e^{j 2 \pi \left( \frac{nk}{N} - \frac{m l}{M} \right)}  {\Bigg [} \sqrt{T} \hspace{-2mm} \sum\limits_{k' = -\infty}^{\infty} \hspace{-2mm} e^{j 2 \pi m \Delta f (\tau + k'T) } \nonumber \\
 & &   \hspace{16.5mm} g(\tau + k'T - nT) e^{-j 2 \pi \nu k' T} {\Bigg ]}
\end{eqnarray}
\normalsize}
where step (a) follows from (\ref{zdef1}) and step (b) follows from substituting the expression 
for $\phi_{k,l}(t)$ from (\ref{zprf61}) into the R.H.S. of step (a). 
The ZAK transform of $g(t - nT)$ is

{\vspace{-4mm}
\label{zprf63}
\begin{eqnarray}
\label{gtmnt}
\sqrt{T} \hspace{-1mm} \sum\limits_{k' = -\infty}^{\infty} \hspace{-2mm} g(\tau + k'T - nT) e^{-j 2 \pi k' \nu T}  \mya e^{- j 2 \pi \nu n T} {\mathcal Z}_g(\tau,\nu) 
\end{eqnarray}
\normalsize}
where ${\mathcal Z}_g(\tau,\nu)$ is the ZAK transform of $g(t)$ and step (a) follows from the first and third equality in (\ref{zprf89}).
Using (\ref{gtmnt}) in (\ref{zprf62}) along with the fact that $e^{j 2 \pi m \Delta f (\tau + k' T)} = e^{j 2 \pi m \Delta f \tau}$ (since $ T \Delta f = 1$)
completes the proof.

\end{document}